# In-Situ Assessment of Array Antenna Currents for Real-Time Impedance Tuning

Charles Baylis[#1], Adam Goad[#2], Trevor Van Hoosier[#3], Austin Egbert[#4], and Robert J. Marks II[#5]
[#]Wireless and Microwave Circuits and Systems Program, Baylor University
[1]Charles_Baylis@baylor.edu, [2]Adam_Goad@baylor.edu, [3]Trevor_VanHoosier@baylor.edu,
[4]Austin_Egbert@baylor.edu, [5]Robert_Marks@baylor.edu

*Abstract*—Impedance tuning has shown promise to maximize output power, gain, or efficiency for transmitter amplifier devices in changing frequency or array environments. Using impedance tuning in phased-array elements, however, has a significant and dynamically changing impact on the transmitted array pattern. To assess the impact of impedance tuning operations on the array pattern, this paper presents a simple method to monitor the antenna input current in real time, allowing optimizations that attempt to preserve the array pattern while achieving goals of increased power, gain, or efficiency. The real-time monitoring of current provides many advantages over traditional array calibration methods in situations where the array element chains can change significantly in magnitude and/or phase during operations.

*Index Terms*— phased arrays, array calibration, radio spectrum management, reconfigurable circuits

## I. INTRODUCTION

Real-time, reconfigurable impedance tuning unleashes the potential of multiple transmitter benefits, such as output power, greater transmission range, and dynamic control over nonlinear distortion artifacts in both spectrum and array pattern [1, 2, 3] for applications where frequency or array transmit angle must change "on the fly". However, the actual impedance tuning operation within an array element during operation adjusts the element transmission, resulting in changes to the array pattern. Because of this, real-time optimizations of element amplifier impedance tuners must also take into account the impact of a tuning operation on the array pattern. Assessments of these significant changes through measurement requires a novel in-situ measurement approach. This method is being designed for application in a fifth-generation (5G) wireless transmitter emulator for real-time coexistence with weather radiometers at 24 GHz.

Existing array calibration approaches require voltage magnitude and phase assessment of each array element. The differences between magnitude and phase are used to provide offsets. For a transmitter, the input signals can be magnitude and phase offset based on calibration results. For a receiver, the magnitude and phase can be adjusted in the signal processing upon receive to correct for these differences. Phased errors resulting from incorrect calibrations, for example, can have significant system-level impact, such as incorrect radar detection and loss of detection accuracy [4]. Pohlmann describes the use of a Bayesian algorithm using Bayesian information and the Bayesian Cramer-Rao Bound to calibrate in a situation where propagation parameters are unknown [5]. Peccarelli presents a method to use mutual-coupling sensing of intermodulation content for array calibration, providing nonlinear equalization of the receiver and digital predistortion of the transmitter in a co-located transmitter/receiver [6]. In-situ array calibration is defined as a measurement that calibrates the array during its actual performance in a fielded system. Takahashi presents measurement of the rotating element electric field vector to assess field amplitudes at selected angles [7]. Sippel uses near-field measurement of a known beacon signal to perform in-situ calibration, which can correct for mutual coupling and multipath [8]. Nicolas presents a method to use a transmitter placed close to the array surface and moving along the array with a tape measure [9]. Salazar shows that a unmanned aerial system (UAS) using an RF probe can be used to measure antenna patterns during array operation [10]. Srinivas demonstrates the calibration of an aircraft-mounted, multifunction array to compensate for fuselage deformation during flight, using both in-situ sources and signals of opportunity [11]. Lebron presents a self-calibration approach using arbitrary amplitude and phase values and evaluating performance with the root mean-squared error [12].

Fulton describes in-situ calibration as an in-field calibration which can either be external, requiring equipment outside of the deployed array structure, or internal, with calibration using equipment internal to the array structure [13]. Using this definition, we present an in-situ, internal adjustment approach that goes beyond simple array calibration. Traditional array calibration approaches become much more difficult when impedance tuning is implemented in real time. A reconfigurable impedance tuner, placed between the amplifier and antenna, causes changes in magnitude and phase each time tuning is performed. If the changes in magnitude and phase are not the same in all elements, then the array pattern can be undesirably altered, as shown by Rodriguez-Garcia [14]. Knowledge of how the array pattern is distorted by an impedance tuning operation informs the optimization of the tradeoffs accomplished through the operation. This information is vital to the real-time tuning optimization algorithms in an array, where synchronization between channels is vital.

Kibaroglu actually demonstrates that array calibration is not needed in some situations where antennas and circuitry utilize printed circuit boards and eliminating connectors [15]. This is because fabrication is precise, and the lack of connectors eliminates a key source of magnitude and phase variability. While this approach works in some applications, real-time understanding of current magnitudes and phases is needed in when reconfigurable impedance tuning is used within the array elements, causing the magnitude and phase of the transmitted current to be altered differently across the array elements. However, this approach would become significantly more elaborate if impedance tuning were used, unless all tuner states were carefully pre-characterized. The in-situ assessment of the antenna input current, discussed in this paper, allows simple monitoring of the array pattern in real-time as part of the optimization process.

## II. IN-SITU ARRAY ASSESSMENT

The proposed approach is different from a traditional array calibration, because it does not require complete knowledge of the transmission parameters of each element, but actually assesses the antenna currents in real time. Many array calibration approaches assess the $S_{21}$ of each element. However, $S_{21}$ is the ratio of the voltage wave entering the antenna to the voltage wave incident on the element from the source. Because antenna surface current actually determines array pattern (and not voltage), it is crucial to measure antenna current in any situation where the impedances presented by the antennas to the array elements are not identical across all elements. This situation occurs when the mutual coupling scenario is not the same for all elements, a situation commonly encountered in practice. Applying real-time impedance tuning adjustments further complicates the analysis, as mutual-coupling scenarios change with impedance tuning adjustments, causing the voltage-current proportions to change in real-time. As such, measurements of the antenna input currents are essential in this operation.

Fig. 1 shows a block diagram of the proposed in-situ current measurement within an array transmitter element. The current entering the antenna ($I_{ant}$) is given in terms of the voltage across the antenna input ports ($V_{ant}$) and the impedance seen looking into the antenna ($Z_{ant}$) as follows:

$$I_{ant} = \frac{V_{ant}}{Z_{ant}} \quad (1)$$

$Z_{ant}$ can be expressed equivalently by a reflection coefficient $\Gamma_{ant}$ with respect to a desired reference impedance $Z_0$:

$$Z_{ant} = Z_0 \frac{1 + \Gamma_{ant}}{1 - \Gamma_{ant}}. \quad (2)$$

$\Gamma_{ant}$ is defined in terms of the forward and reverse traveling waves at the antenna with respect to the same reference impedance:

$$\Gamma_{ant} = \frac{V_{ant}^-}{V_{ant}^+} \quad (3)$$

$V_{ant}$ can be expressed as a sum of voltage waves entering and leaving the antenna, respectively:

$$V_{ant} = V_{ant}^+ + V_{ant}^- = V_{ant}^+ + \Gamma_{ant} V_{ant}^+ \quad (4)$$

Substituting (2) and (4) into (1) gives

$$I_{ant} = \frac{V_{ant}^+ + \Gamma_{ant} V_{ant}^+}{Z_0 \frac{1 + \Gamma_{ant}}{1 - \Gamma_{ant}}} \quad (5)$$

Substituting (3) gives

$$I_{ant} = \frac{V_{ant}^+ + \frac{V_{ant}^-}{V_{ant}^+} V_{ant}^+}{Z_0 \frac{1 + \frac{V_{ant}^-}{V_{ant}^+}}{1 - \frac{V_{ant}^-}{V_{ant}^+}}} = V_{ant}^+ \frac{1 + \frac{V_{ant}^-}{V_{ant}^+}}{Z_0 \frac{1 + \frac{V_{ant}^-}{V_{ant}^+}}{1 - \frac{V_{ant}^-}{V_{ant}^+}}} = \frac{V_{ant}^+ - V_{ant}^-}{Z_0} \quad (6)$$

Equation (6) indicates that the current entering the antenna can be calculated by measurement of the incident and return voltage waves from the antenna. Fig. 1 shows a measurement setup for these voltages using a dual-directional coupler. A voltage measurement instrument, such as a software defined radio, can be used to monitor the forward and reverse traveling waves.

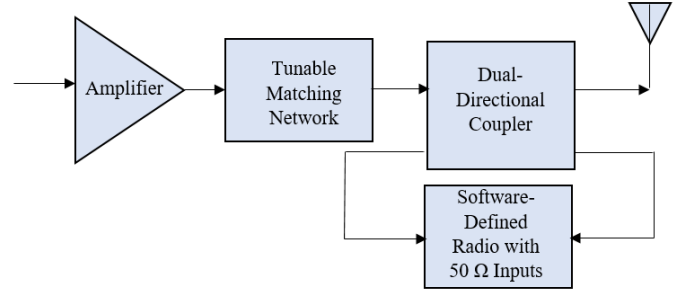

Fig. 1. In-situ current measurement in a transmitter array element with a tunable matching network

The ability to calculate the currents entering the array element antennas allows direct calculation of the array pattern. The calculation of the array pattern can be performed upon impedance tuning optimization measurements to indicate the impact of an impedance tuning operation on the transmitted array pattern. Having the simple capability to perform dynamic, in-situ measurement of the antenna currents is crucial in an array with real-time optimizable circuits.

There are additional benefits to this approach, as compared to a potential alternative path of implementing traditional array calibration. Traditional array calibration measures the transmission coefficient vectors of the array elements, providing the ability to adjust the input voltages to get the desired output voltages. Because the voltage transmission coefficients are used, traditional array calibration relies on voltage measurements at the antennas, rather than current measurements. The assessment of the array pattern through the voltages across the element antennas is accurate only if the impedances presented by all antennas are the same. Typically, mutual coupling effects keep the antenna impedances different. While advanced approaches such as those described by Takahashi [7] and Sippel [8] can include mutual coupling effects, traditional array calibration is an approach of bookkeeping and correction rather than simple output monitoring. The measurement of the current we describe will remove the need for extensive array calibration and re-performance of in-situ array calibration. To use array calibration with impedance tuning, the S-parameters of the impedance tuner would be needed at every magnitude and phase to update the transmission characteristics of each element and understand the transmitted array pattern.

The current-monitoring approach we propose is relatively hands-off compared to traditional array calibration. Our approach does not require an end-to-end $S_{21}$ measurement, knowledge of any of the array elements, or impedance tuner characterization. Special approaches for adjustment, based on mutual coupling or other impedance changes involving the array elements, are not needed. The downside is the additional hardware required: a dual-directional coupler and capability to monitor two voltages within each array element.

### III. SIMULATION RESULTS

To validate this method, a simple simulation was performed in Keysight Advanced Design System (ADS) using the schematic shown in Fig. 2. The dual-directional coupler used is modeled after the typical parameters of the RF-Lambda RFDDC8G26G15 coupler.

The coupled ports were terminated in 50 Ω to represent the termination of the measurement device. In this simulation the total voltage assessed on the nodes labeled V_P3 and V_P4 represent the measurements that will be made in practice. To determine the forward and reverse travailing waves the scattering effects of the coupler must be de-embedded. An S-parameter simulation is performed simultaneously on an identical coupler to provide the characterization needed to de-embed back through the coupler. The modeled coupler has 24 GHz S-parameters, with respect to $Z_0 = 50$ Ω, of

$$S = \begin{bmatrix} .130\angle 0° & .859\angle 0° & .178\angle 90° & .032\angle 0° \\ .859\angle 0° & .130\angle 0° & .032\angle 0° & .178\angle 90° \\ .178\angle 90° & .032\angle 0° & .130\angle 0° & .000\angle 0° \\ .032\angle 0° & .178\angle 90° & .000\angle 0° & .130\angle 0° \end{bmatrix}.$$

After finding $V_{ant}^+$ and $V_{ant}^-$, $I_{ant}$ is calculated using equation (6). These are compared with direct assessment of the current using the Current Probe tool in ADS in Table 1.

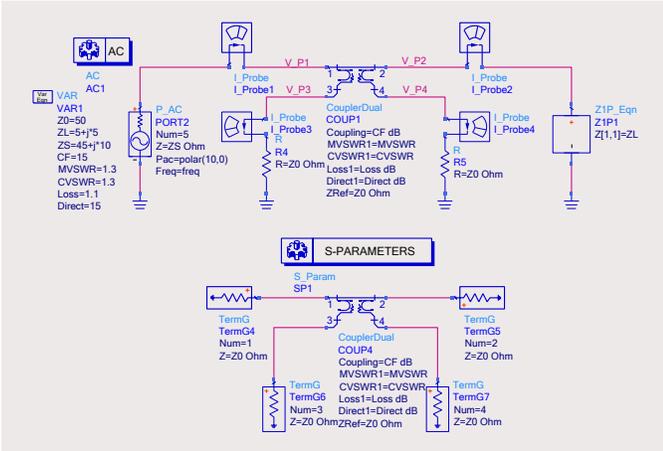

Fig. 2. ADS simulation schematic for method validation

With respect to the coupler ports as numbered in Fig. 2, if the four-port S-parameters of the coupler are known, then $V_{ant}^+$ and $V_{ant}^-$ can be calculated using the known values of $V_3^-$ and $V_4^-$. In this situation, the antenna is connected to port 2 of the coupler, meaning that $V_{ant}^+ = V_2^-$ and $V_{ant}^- = V_2^+$ (the wave entering the antenna is the wave leaving port 2 of the coupler, and the wave leaving the antenna is the wave entering port 2 of the coupler. The equations describing the four-port S-parameters of the coupler are as follows:

$$V_1^- = S_{11}V_1^+ + S_{12}V_2^+ + S_{13}V_3^+ + S_{14}V_4^+ \qquad (7)$$

$$V_2^- = S_{21}V_1^+ + S_{22}V_2^+ + S_{23}V_3^+ + S_{24}V_4^+ \qquad (8)$$
$$V_3^- = S_{31}V_1^+ + S_{32}V_2^+ + S_{33}V_3^+ + S_{34}V_4^+ \qquad (9)$$
$$V_4^- = S_{41}V_1^+ + S_{42}V_2^+ + S_{43}V_3^+ + S_{44}V_4^+ \qquad (10)$$

In this simulation, the terminations of the coupler on ports 3 and 4 are $Z_0 = 50$ Ω. As such, no wave travels inward at either of these ports, and $V_3^+ = V_4^+ = 0$. This simplifies equations (7) through (10) to the following:

$$V_1^- = S_{11}V_1^+ + S_{12}V_2^+ \qquad (11)$$
$$V_2^- = S_{21}V_1^+ + S_{22}V_2^+ \qquad (12)$$
$$V_3^- = S_{31}V_1^+ + S_{32}V_2^+ \qquad (13)$$
$$V_4^- = S_{41}V_1^+ + S_{42}V_2^+ \qquad (14)$$

In this case, the measured quantities available are $V_3^-$ and $V_4^-$, and it is desired to calculate $V_2^+ = V_{ant}^-$ and $V_2^- = V_{ant}^+$. Equation (13) can be solved for $V_1^+$ in terms of $V_3^-$ and $V_2^+$:

$$V_1^+ = \frac{V_3^- - S_{32}V_2^+}{S_{31}}. \qquad (15)$$

This is substituted into equation (14) to solve for $V_2^+$ in terms of $V_3^-$ and $V_4^-$:

$$V_4^- = S_{41}\left(\frac{V_3^- - S_{32}V_2^+}{S_{31}}\right) + S_{42}V_2^+ \qquad (16)$$

Solving this expression for $V_2^+ = V_{ant}^-$ in terms of the measured $V_3^-$ and $V_4^-$ gives

$$V_{ant}^- = V_2^+ = \frac{V_4^- - \frac{S_{41}}{S_{31}}V_3^-}{S_{42} - \frac{S_{41}S_{32}}{S_{31}}}. \qquad (17)$$

Equations (15) and (17) can then be substituted into equation (12) to give $V_2^- = V_{ant}^+$:

$$V_{ant}^+ = V_2^- = S_{21}\left(\frac{V_3^- - S_{32}V_2^+}{S_{31}}\right) + S_{22}V_2^+ \qquad (18)$$

$$V_{ant}^+ = V_2^- = \frac{S_{21}}{S_{31}}V_3^- + \left(S_{22} - \frac{S_{21}S_{32}}{S_{31}}\right)V_2^+ \qquad (19)$$

$$V_{ant}^+ = V_2^- = \frac{S_{21}}{S_{31}}V_3^- + \left(S_{22} - \frac{S_{21}S_{32}}{S_{31}}\right)\frac{V_4^- - \frac{S_{41}}{S_{31}}V_3^-}{S_{42} - \frac{S_{41}S_{32}}{S_{31}}} \qquad (20)$$

Using the S-parameters with equations (17) and (20), the desired $V_{ant}^-$ and $V_{ant}^+$ can be calculated from the measured $V_3^-$ and $V_4^-$. $V_{ant}^-$ and $V_{ant}^+$ can then be used in equation (6) to calculate the antenna current $I_{ant}$. This allows the array pattern to be calculated.

This method for current assessment is performed using the simulation values of voltages V_P3 and V_P4 from Fig. 2. These values are actually the total voltages $V_3$ and $V_4$ from the equations, respectively, but because ports 3 and 4 are terminated in $Z_0$, $V_3 = V_3^-$ and $V_4 = V_4^-$.

Table I shows the results of the validation simulation exercise. Five tests were performed with different impedance terminating the coupler on its port 1 and port 2 ($Z_s$ and $Z_L$, respectively). For each of these tests, different values were measured for $V_3$ (labeled as V_P3 in Fig. 2) and $V_4$ (labeled as V_P4 in Fig. 2). For each case, the current calculated using equations (6), (17), and (20) ($I_{calc}$) is exactly the same as the current measured using the "IProbe_2" element in Fig. 2. This validates the capability of using these voltages with a dual-directional coupler to calculate the antenna current.

TABLE I: SIMULATION VALIDATION TEST RESULTS

| # | $Z_s(\Omega)$ | $Z_L(\Omega)$ | $V_3(V)$ | $V_4(V)$ | $I_{meas}(A)$ | $I_{calc}(A)$ |
|---|---|---|---|---|---|---|
| 1 | 50 | 50 | 5.62/90° | 1.00/90° | 0.54/0° | 0.54/0° |
| 2 | 50 | 5 | 5.66/96° | 3.71/-74° | 0.89/0° | 0.89/0° |
| 3 | 50 | 5+$j$15 | 5.98/95° | 3.29/-105° | 0.87/-12° | 0.87/-12° |
| 4 | 15 | 50 | 4.43/90° | 0.787/0° | 0.43/0° | 0.43/0° |
| 5 | 15+$j$8 | 50 | 5.58/75° | 0.783/-6° | 0.43/-6° | 0.43/-6° |

## IV. CONCLUSIONS

A method of measuring the current entering the antenna of a phased-array element using coupled voltage measurements with a dual-directional coupler has been described and validated using simulations. This method can be used in situations where magnitude and phase of element transmissions varies in real time, such as when reconfigurable impedance tuners are placed in the array elements. Importantly, this approach may be used in place of traditional array calibration for some applications, because it is capable of directly assessing the antenna currents, which can be used to calculate the array pattern. The use of this approach, instead of having to re-perform array calibrations for all dynamic array element settings, in situations where the array element magnitudes and phases are dynamically changing (such as in real-time impedance tuning) is expected to be especially convenient.

## ACKNOWLEDGMENTS

This research has been funded by the National Science Foundation (Grant No. 2030243).

## REFERENCES

[1] J. Alcala-Medel, A. Egbert, C. Calabrese, A. Dockendorf, C. Baylis, G. Shaffer, A. Semnani, D. Peroulis, A. Martone, E. Viveiros, and R.J. Marks II, "Fast Frequency-Agile Real-Time Optimization of High-Power Tuning Network for Cognitive Radar Applications," 2019 IEEE MTT-S International Microwave Symposium, Boston, Massachusetts, June 2019.

[2] M. Fellows, C. Baylis, L. Cohen, and R.J. Marks II, "Real-Time Load Impedance Optimization for Radar Spectral Mask Compliance and Power Efficiency," *IEEE Transactions on Aerospace and Electronic Systems*, Vol. 51, No. 1, January 2015, pp. 591-599.

[3] P. Rodriguez-Garcia, J. Sifri, C. Calabrese, A. Goad, C. Baylis, and R.J. Marks II, "Spurious Beam Suppression in Dual-Beam Phased Array Transmission by Impedance Tuning," accepted for publication in *IEEE Transactions on Aerospace and Electronic Systems*, December 2021.

[4] P.J. Flament, M. Flament, C. Chavanne, X. Flores-Vidal, I. Rodriguez, L. Marie, and T. Hilmer, "In-situ Calibration Methods for Phased Array High Frequency Radars," American Geophysical Union, Fall Meeting 2016, Abstract #OS13B-1824, December 2016.

[5] R. Pohlmann, S Zhang, E. Staudinger, S. Caizzone, A. Dammann, and P.A. Hoeher, "Bayesian In-Situ Calibration of Multiport Antennas for DoA Estimation: Theory and Measurements," *IEEE Access*, Vol. 10, April 2022, pp. 37967-37983.

[6] N. Peccarelli and C. Fulton, "A Mutual Coupling Approach to Digital Pre-Distortion and Nonlinear Equalization Calibration for Phased Arrays," 2019 IEEE International Symposium on Phased Array System and Technology," Waltham, Massachusetts, October 2019.

[7] T. Takahashi, Y. Koniski, and I. Chiba, "A Novel Amplitude-Only Measurement Method to Determine Element Fields in Phased Arrays," *IEEE Transactions on Antennas and Propagation*, Vol. 60, No. 7, July 2012, pp. 3222-3230.

[8] E. Sippel, M. Lipka, J. Geiβ, M. Hen, and M. Vossiek, "In-Situ Calibration of Antenna Arrays Within Wireless Locating Systems," *IEEE Transactions on Antennas and Propagation*, Vol. 68, No. 4, April 2020, pp. 2832-2841.

[9] J.-J. Nicolas, "Measurements of Phased Array Antenna Fields In Situ: A Few Key Aspects," 2008 IEEE Radar Conference, Rome, Italy, May 2008.

[10] J.L. Salazar, A. Umeyama, S. Duthoit, and C. Fulton, "UAS-Based Antenna Pattern Measurements and Radar Characterization," 2018 IEEE Conference on Antenna Measurements and Applications," Sweden, 2018.

[11] S. Srinivas and D.W. Bliss, "Conformal Multi-Service Antenna Arrays: Hybrid In Situ & Signal of Opportunity (SoOP) Calibration," 2020 IEEE 92nd Annual Vehicular Technology Conference, Victoria, British Columbia, Canada, November 2020.

[12] R.M. Lebron, P.-S. Tsai, J.M. Emmett, C. Fulton, and J.L. Salazar-Cerreno, "Validation and Testing of Initial and In-Situ Mutual Coupling-Based Calibration of a Dual-Polarized Active Phased Array Antenna," *IEEE Access*, Vol. 8, March 2020, pp. 78315-78329.

[13] C. Fulton, M. Yeary, D. Thompson, J. Lake, and A. Mitchell, "Digital Phased Arrays: Challenges and Opportunities," *Proceedings of the IEEE*, Vol. 104, No. 3, February 2016, pp. 487-503.

[14] P. Rodriguez-Garcia, J. Sifri, C. Calabrese, C. Baylis, and R.J. Marks II, "Range Improvement in Single-Beam Phased Array Radars by Amplifier Impedance Tuning," 2021 IEEE Texas Symposium on Wireless and Microwave Circuits and Systems, Waco, Texas, May 2021.

[15] K. Kibaroglu, M. Sayginer, T. Phelps, and G.M. Rebeiz, "A 64-Element, 28-GHz Phased-Array Transceiver With 52-dBm EIRP and 8-12 Gb/s 5G Link at 300 Meters Without Any Calibration," *IEEE Transactions on Microwave Theory and Techniques*, Vol. 66, No. 12, December 2018, pp. 5796-5811.